# How to Measure Sessions of Mobile Device Use? Quantification, Evaluation, and Applications[*]

Jonathan J. H. Zhu[1,†], Hexin Chen[1], Tai-Quan Peng[2], Xiao Fan Liu[3], and Haixing Dai[1]

[1]Web Mining Lab, Dept. of Media & Communication, City University of Hong Kong
[2]Dept. of Communication, Michigan State University, USA
[3]School of Computer Science and Engineering, Southeast University, China

**Abstract**

Research on mobile phone use often starts with a question of "How much time users spend on using their phones?". The question involves an equal-length measure that captures the *duration* of mobile phone use but does not tackle the other temporal characteristics of user behavior, such as *frequency*, *timing*, and *sequence*. In the study, we proposed a variable-length measure called "session" to uncover the unmeasured temporal characteristics. We use an open source data to demonstrate how to quantify sessions, aggregate the sessions to higher units of analysis within and across users, evaluate the results, and apply the measure for theoretical or practical purposes.

Research on mobile phone use often starts with a question of "How often users use their phones?". Although seemingly simple and straightforward, this question involves at least two different dimensions: how much time (e.g., minutes) per given unit (e.g., a day) versus how many times (i.e., frequency) per given unit. The former is based on an *equal*-length scale (each minute being the same length) whereas the latter on a *variable*-length scale (each occurrence lasting differently).

Fig. 1 illustrates the similarities and differences between equal- and variable-length measures for the same temporal behavior of a user (ID = 667), randomly selected from our dataset (to be described later). For simplicity, we show only 3 days of his records. Fig. 1a plots the number of seconds the user spent on his phone, whereas Fig. 1b shows the length of each task (in black) and the length of idle time between two tasks (in white). As such, Fig. 1a highlights the *duration* of task time along the Y-axis, whereas the Fig. 1b describes the *frequency*, *timing,* and *sequence* of task time along the X-axis.

Figure 1. Equal- and Variable-length Measures of Mobile Use Time[*]

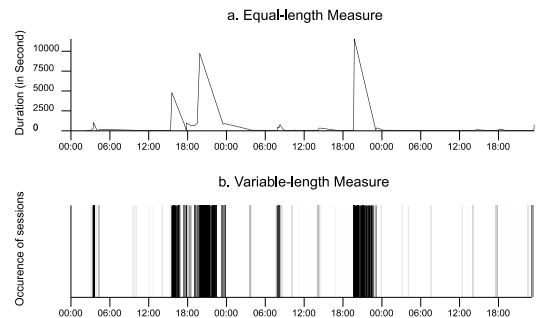

[*] The user is randomly selected from the Cambridge data, with only 3 consecutive days shown for illustration purpose. The number of sessions per day for the user is around the median level of the sample.

Time data based on equal-length measures have several desirable properties. Equal-length measures are of high face validity because they are directly observable without any data transformation. As a natural scale, they are straightforwardly intuitive for presentation and comprehension. Therefore, equal-length measures are the most popular way to quantify media use, either traditional media (e.g., TV or newspapers) or new media (web or mobile phone). For instance, the use of mobile phones is typically quantified as the total amount of time per day (or hour, week, etc.) spent on applications ("apps" hereafter) (Boase & Ling, 2013). Other scholars have adopted innovative ways (e.g., experience sampling) to record the media use time in the context of multitasking (David, Srivastava, & Kim, 2013; Wang & Tchernev, 2012). However, their measures of time (e.g., behavior duration) are of an equal-length scale.

However, equal-length measures cannot adequately capture all the temporal characteristics of the mobile use behavior. User time involves at least four dimensions: (1) duration (amount of time), (2) frequency (number of tasks), (3) timing (start and

---

[*] The study was supported in part by CLASS GWF (9618003) and SRG (7004513) of City University of Hong Kong.
[†] Corresponding author, email address: j.zhu@cityu.edu.hk

Preprint for forthcoming in *Mobile Media & Communication.*

end of each task), and (4) sequence (flow of adjacent tasks). Equal-length measures are appropriate to quantify duration. However, these measures provide limited information about frequency, timing, or sequence. For instance, if 3 users each spent 120 minutes a day on mobile phones but acted differently in terms of how to allocate the time. User A concentrated the use in the morning, user B in the evening, and user C throughout the day. Their duration of use will look identical although they followed distinct temporal patterns, e.g., A and B have different timing, and C has a high frequency. Such differences across users can be easily quantified by an alternative measure called "session".

Conceptually, session is a continuous uninterrupted sequence of tasks (e.g., watching, listening, and interacting) a user performs on a media device. As such, the length of sessions varies not only across users but also within the same user at different times of the day and on different days. Operationally, session is an ordinal variable (indexed by an ID) that is associated with a set of variables to describe the properties of the session, including the start time, end time, duration, inter-session time, and content. Therefore, a session captures the full range of the temporal characteristics of user behavior, which can be used to address many unsolved theoretical or practical questions about mobile phone use. For instance, a longstanding question is whether the use of mobile phones has caused an increasing fragmentation on the workflow or face-to-face socializations of people (Karlson et al., 2010). The hypothesis can be tested directly on the basis of changes in the frequency of sessions over time. Likewise, session provides informative insights for mobile application design, mobile advertising scheduling, and other mobile communication practices.

**Previous Studies on Session**

*Duration of TV Viewing*

Variable-length measures have been used to measure various media use behaviors, including viewing TV programs, browsing webpages, and using mobile phones. In the research on TV viewing, "viewing duration" or "time span" is used to quantify viewing sequences (Webster, 1985). These measures introduce a series of theoretical concepts about audience temporal patterns, such as repeated viewing (on the same program at the same time across different days), audience duplication (across adjacent programs on the same day), and channel switching (across different channels during the adjacent time segments). These concepts are directly applicable to mobile phone sessions if "TV programs" are replaced with "mobile apps".

*Session in Web Browsing*

Previous studies on web browsing behavior have used the term "session" to describe the set of consecutive and related visits to webpages initiated by users. A session is defined as a variable-length measure. A threshold-based approach is widely used to identify sessions, which assumes the continuity of user browsing behavior. Therefore, a substantially long break between two adjacent requests is considered evidence of the expiration of a session (Mehrzadi & Feitelson, 2012). Certain studies have adopted a global threshold (e.g., 30 minutes) to define the break between sessions for all users (Arlitt, 2000; Menasc et al., 1999). The global threshold makes a strong (but generally unrealistic) assumption about the homogeneity of users. Other studies have adopted a user-specific threshold to extract sessions for each user (Mehrzadi & Feitelson, 2012; Murray, Lin, & Chowdhury, 2006; Ware, Page, & Nelson, 1998).

The extraction of browsing sessions provides useful insights to measure mobile phone use. Websites and mobile phones are all computer systems with which users interact. The systems record user-initiated tasks on a fine-grained temporal unit. Nevertheless, user behavior on mobile phones is much more complicated than that on websites because mobile phones integrate many social and business functions to become a fixture of our daily lives (Ling, 2004).

*Session in Mobile Phone Use*

Previous studies have used two ways to measure sessions of mobile phone use: *threshold*-based versus *screen*-based. The threshold approach is inherited from web browsing research that has defined a session based on a threshold determined by the duration of inactiveness, such as 30 seconds in Böhmer et al. (2011) or 10 seconds in Van Canneyt et al. (2017). The threshold is arbitrarily chosen. In addition, the approach does not distinguish user-initiated tasks from machine-operated tasks.

The screen-based approach has been adopted to overcome these problems. The approach defines a session based on the deactivation of the screen of the phone, assuming that the screen status is a valid indicator of intentional human behavior. Falaki et al. (2010) treated the duration whenever the screen is on, a voice call is active, or an app runs in the foreground as a session. Yan et al. (2012) followed the same logic by defining a session as a sequence of apps launched between the unlocking and relocking of the screen.

Although most of existing studies have aimed to find global regularities underlying the user behavior on mobile phone uses, noting that certain studies (Falaki et al., 2010) have explicitly acknowledged the existence of individual differences. Simultaneously uncovering both global regularities and individual variabilities is desirable.

**Research Questions**

By considering the variety of approaches in the literature, we define session as a continuous sequence of tasks initiated by a user on a media device. The term "sequence of tasks" indicates that we have adopted the multi-app version of session. The emphasis on "user initiation" excludes tasks that are activated by machine (see more discussions later). "Media device" exclusively refers to mobile phones in the current study. However, the definition should also be applicable to other media devices whether they are old or new, fixed or mobile.

Given the methodological and exploratory nature of the current study, we do not impose any theoretically driven hypothesis. Instead, we organize our data collection, analysis, and presentation around the following questions:

1. How to construct and quantify sessions from massive (and commonly noisy) mobile phone logs?

2. How to evaluate the validity of the resulting sessions and session-based measures?

3. How to apply the session-based measures to theoretical context (e.g., hypothesis-testing) and practical context (e.g., user profiling)?

**Method**

*Data*

Our data were obtained from an open source provided by the Device Analyzer project at Cambridge University ("Cambridge data" hereafter) that has used an app to collect mobile phone logs from volunteer participants (Wagner, Rice, & Beresford, 2013). Over 31,000 users installed the app on their Android-based phones between December 2010 and February 2016. Of the users in the current study, 4,017 are "active" and have valid records on 10+ days (median = 130 days).

*Session Construction*

Similar to other mobile phone logs, the Cambridge data contain the start and end time of phone tasks (e.g., phone calls, short messages, and apps) in milliseconds. However, two technical problems in the dataset hinder the straightforward extraction of sessions. The first problem is that many log records are machine-activated by either an operating system or apps. Thus, these tasks should not be counted as session time. Moreover, the dataset has no information that indicates whether tasks are operated by machine or by human. We have adopted a screen-based approach to address the problem, i.e., treating screen unlocking and locking as the start and end time of a human-operated session, respectively (Fig. 2).

Figure 2. Hypothetical "Session" of Mobile Phone Use

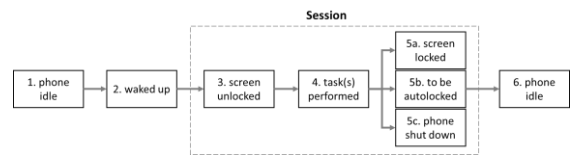

The second problem involves the situations where multiple apps are simultaneously running in the foreground, which defies the logic that only one foregrounded app should be running at any given time. Whether this is caused by an idiosyncratic method used by the Cambridge project or is generally applicable to (some versions of) Android phones remains unknown. As such, the problem makes the measurement of a single-app session impossible (Van Canneyt, Bron, & Haines, 2017). In our approach, when a user performs several tasks (talking, texting, and using apps) between the unlocking and locking of the screen, we treat the multiple tasks as a continuous session (i.e., a multi-app session).

We develop an algorithm to identify and construct sessions from the log data. For each user, the algorithm iterates through each line of logs to capture the wake-up time, unlocked time, and locked/shutdown time of the phone. The algorithm then constructs sessions based on the unlocking and locking time (see the pseudo code in Appendix 1).

The resulting dataset (called "sessions") contains each session as a row, with user ID, session ID, start time, end time in the columns, and multiple rows per user. The structured data are directly applicable for certain purposes (e.g., testing fragmentation trend shown later), but can be excessively fine-grained and noisy for many other purposes. Therefore, we perform the following steps to aggregate sessions to help uncover the temporal patterns within and across users.

*Community Detection of Sessions*

Time has a built-in hierarchical structure that can be leveraged to characterize temporal patterns. In this study, we aggregated sessions successively into two higher levels of granularity: (i) *clusters* of similar sessions per user; and (ii) *communities* of similar users of the entire sample. We used network community detection method to implement the aggregation. Specifically, we choose the Louvain algorithm (Blondel et al., 2008) because of its computing efficiency and effectiveness (e.g., resulting in a community structure with a higher modularity than do other algorithms). Appendix 2 presents the technical details of the community detection.

The first aggregation (from sessions to session-clusters) inputs a "sessions" dataset and then outputs a "session-clusters" dataset that contains each session-cluster as a row (i.e., the unit of analysis), user ID, cluster ID, N of sessions in the cluster, cluster modularity (measuring the quality of the cluster), and centroid (measuring the geometric mean) as variables. The second aggregation (from session-clusters to user-communities) inputs a session-clusters dataset and then outputs a "user-communities" dataset that contains each user as a row (the unit of analysis), user ID, community-ID, and N of clusters in the community as variables.

**Results**

*Sessions*

Table 1 summarizes the descriptive statistics of mobile phone use in equal-length measure (the amount of time) and variable-length measure (the number of sessions).

Table 1. Descriptive Statistics of Sessions per User per Day[*]

|  | 1. Total Length of Mobile Time (in min.) | 2. N of Sessions | 3. Mean Length of Mobile Time (in min.) |
|---|---|---|---|
| Mean | 175 | 24 | 17 |
| Median | 97 | 18 | 4 |
| Minimum | 0.1 | 1 | 0.1 |
| Maximum | 2,680 | 466 | 1,440 |

[*] The unit of analysis is user-day-session (i.e., sessions per day per user); N of total users = 4,017; N of total sessions = 18.2 million.

*Amount of time.* Each user spends an average of 175 minutes (or about 3 hours) on the phone per day (col. 1). However, the amount of time is highly-uneven across users, following a power-law distribution (Fig. 3a) that has been widely observed (Candia et al., 2008). Therefore, it will be appropriate to describe the length of time in the median (= 97 minutes or 1.5 hours, col. 1).

Figure 3. Distribution of Equal- and Variable-length Measures[*]

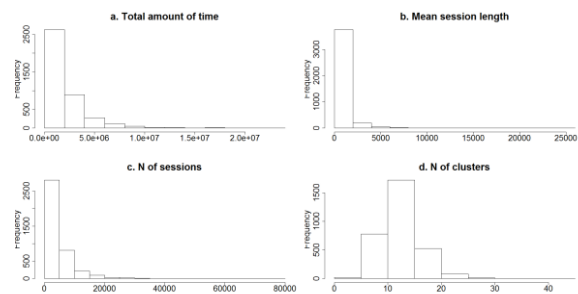

[*] All the measures are based on per user per day for a–c (i.e., a shows the total length of time per user per day, b shows the mean length of sessions per user per day, and c shows the number of sessions per user per day), and d shows the number of clusters per user.

*Number of sessions.* A total of 18.2 million sessions are identified from all users. Each user engages, on average, in 24 sessions per day (median = 18, col. 2). In other words, on average, users pick up their phones about 20 times a day. Each session lasts, on average, 17 minutes (median = 4 minutes, col. 3). The mean length and the number of sessions also follow a power-law distribution (Fig. 3b and 3c, respectively), which means that most users use their phone only a few times per day (each for a short duration), whereas a few others engage in an excessive number of times per day (each for a lengthy period). Altogether, the results show that the users in the dataset use their phone actively but not excessively.

*Session-Clusters*

We aggregated 4,533 (median = 3,008) sessions per user to 13 (median = 12, col. 1 of Table 2) session-clusters per user. Each session-cluster contains an average of 339 (median = 245) sessions (col. 2). The resulting session-clusters provide an easy way to detect, conceptually and visually, daily rhythms of mobile phone use.

In addition, the number of session-clusters approximates a normal distribution (Fig. 3d), which differs sharply from the power-law distribution observed at the session level (Figs. 3a–3c). The aggregation of sessions to clusters helps meet the requirement for normality by classic statistical analyses.

Table 2. Descriptive Statistics of Session-Clusters per User for all Days[*]

|  | 1. N of Session-Clusters | 2. Size of Session-Clusters | | 3. Modularity of Clustering Structure |
|---|---|---|---|---|
|  |  | Mean Size | Median Size |  |
| Mean | 13 | 339 | 369 | 0.77 |
| Median | 12 | 242 | 245 | 0.78 |
| Minimum | 1 | 2 | 2 | 0.00 |
| Maximum | 43 | 6,447 | 6,285 | 0.89 |

[*] The unit of analysis is user-cluster-session; N of total users = 4,017; N of total clusters = 51,636; N of total sessions = 18.2 million

*User-Communities*

We identify 13 communities of users from the sample based on the session-clusters of each user. The results are described in the Application section to avoid redundant presentations.

**Evaluation**

We empirically examine whether (i) the new measure produces systematic biases (e.g., consistently over-estimating or under-estimating the targeted phenomenon) and (ii) the new measure is substantially correlated with certain existing measure(s).

*External Validity*

We first compare three key estimates of mobile phone use (i.e., total length, session length, and frequency of sessions, from Tables 1 and 2) with those reported by relevant studies to ascertain the external validity. Conducting a significance test of the comparison is difficult for two reasons: (i) the five studies under comparison are heterogeneous in many aspects and (ii) the required information (e.g., standard deviations) for the significance test is unavailable. The informal comparison is only indicative of the direction and range of observed differences.

Table 3 shows that no systematic difference is observed between our results and benchmarks. Our total length of time is between the findings of Winnick and that of Yang et al. Our mean length of sessions is higher than the findings of Falaki et al. or Van Canneyt but similar to that of Yang et al. Our number of sessions is lower than that of Winnick or Falaki et al. The reason why we obtained fewer sessions but longer session length than what Falaki et al. obtained may be due to the different definitions of session (multiple apps in ours but a single app in Falaki et al.). However, the work of Winnick is based on a multi-app definition but still found 3 times more sessions than what we found. Conversely, Yang et al. used a single-app definition but found a similar mean length of sessions than what we obtained.

Table 3. Comparison with Benchmark Studies[*]

|  | The Current Study | Winnick (2016) | Falaki et al. (2011) | Yang et al. (2015) | Van Canneyt et al. (2017) |
|---|---|---|---|---|---|
| Total Time Length (min.) | 175 (median = 97) | 145 | 30–500 | 300 | n.a. |
| Mean Session Length (min.) | 17 (median = 4) | n.a. | 0.2–4.2 | 8–41 | 5-7 |
| Mean N of Sessions | 24 (median = 18) | 76 | 10–200 | n.a. | n.a. |
| N of Apps per Session | Multiple | Multiple | Single | Single | Single |

[*] The unit of analysis is per user per day for all studies. Some studies only reported results in ranges.

*Discriminant Validity*

When advocating session as a complement to the prevailing equal-length measures of time, we assume that the two measures are distinctive to each other, i.e., discriminant validity between the two. We conduct a multilevel regression with the amount of time per user per day as the dependent variable, the number of sessions per user per day as the independent variable, and the number of active days for the user as a control variable (a weighting factor). Because the distribution of both dependent and independent variables are highly skewed (Figs. 3a and 3c), we perform the regression analysis twice, one based on the original scores and the other based on the log-transformed scores.

The resulting squared semi-partial correlation between the dependent and independent variables is 0.018 (in original scores) or 0.284 (logged). Hence, the degree of redundancy between the number of sessions and the length of time is weak (2%) or modest (28%), depending on the data transformation method. In short, session provides additional information about mobile phone use over and above the mere quantity of time.

*Quality of Clusters and Communities*

We use modularity scores (Newman, 2006) to evaluate the quality of resulting session-clusters and user-communities. Modularity score ranges from 0 to 1, with a higher value indicating a higher likelihood to divide a network to a set of clusters or communities that are internally coherent but are externally distinctive. Col. 3 of Table 2 depicts that the resulting modularity scores for session-clusters are consistently high, varying between 0.77 and 0.80, suggesting that it is appropriate to aggregate individual sessions to coherent clusters. Note that 22,000+ sessions (which are only 0.12% of the total sessions) cannot be aggregated to any cluster because they drastically deviate from the daily rhythm of relevant users.

The resulting modularity score (0.83) for user-communities (see Applications below) is also satisfactory. All users are partitioned into 13 communities, with a compatible membership size for 11 communities (Table 5). Therefore, two exceptionally small communities (#12 and #13, each with only 2 users) are removed from further analysis.

**Applications**

We have applied the session-based measures to three cases, ranging from testing theoretical hypothesis to classifying users. Equal-length measures, such as the amount of time, are either unable or cumbersome to handle these applications.

*Testing Fragmentation Trend Effects*

The hypothesis holds that the use of mobile phone has interrupted the daily life of users into increasingly short segments (Karlson et al., 2010). Although intuitively convincing, the hypothesis has never been empirically verified. The key challenge lies on the type of evidence required: fragmentation is not about the amount of time spent on mobile phones, but about the *pattern* (i.e., frequency and sequence) of the usage. The latter can be adequately measured by session proposed in the study. Specifically, the number of sessions per user per day quantifies the frequency of mobile phone use or the interruption of other daily activities. Furthermore, a growing trend in the number of sessions over time is a valid indicator of the fragmentation trend.

We use the Cambridge data for the test. To ensure reliable estimates, we select users who had records on 10+ days per month for 3 consecutive months. Over 3,100 users meet the criteria. For each user, we test the 3-month trend by regressing the number of sessions per day on the calendar day. Only 19% of users increased the number of sessions over time, whereas 38% had fewer sessions, and 47% showed no significant change during the study period (Fig. 4). Although a detailed interpretation of results goes beyond the scope of the current study, the evidence is neither strong nor consistent in supporting the fragmentation trend hypothesis, at least among the 3,100 users under study. Thus, a longer time window is necessary to test the fragmentation trend.

Figure 4. Changes in the Significance and Direction of Fragmentation Trend

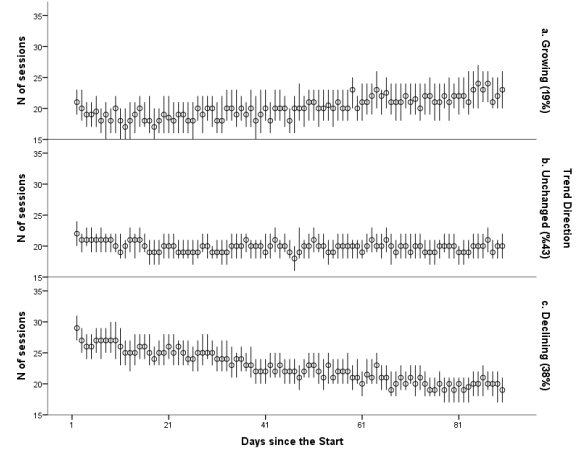

*Quantifying Temporal Regularities*

Temporal regularities of mobile phone users are important but difficult to detect. TV audience researchers use Repeated Viewing to quantify the loyalty to a specific content by measuring how often viewers watch the same program at the same time slot across different days (Webster et al., 2006). Mobile phone use differs from TV viewing in many aspects. However, repetitive tasks on mobile phones signal certain regularities (e.g., habitual behavior, and elastic lifestyle). We use the community detection method to reduce a large number of sessions per user into a small set of clusters for parsimony. Consequently, we obtain session-clusters that are a handy tool to facilitate the identification of the repetitive behavior of mobile phone use.

Specifically, we define a "rate of repeated sessions" (RRS) as follows,

$$RRS_i = \sum_{j=1}^{J} \frac{n_{ij}}{J_i N_i}, \quad (1)$$

where $n_{ij}$ is the number of "session-days" (on which user $i$ has one or more sessions falling to session-cluster $j$ (= 1 to $J$), $N_i$ is the total number of days user $i$ has usage data, and $J_i$ is the total number of session-clusters $i$ has. RRS varies from 0 to 1, with 1 showing that all session-days are covered by a session-cluster. As such, RRS describes the degree of temporal regularity for user *i* to use mobile phone at fixed time slots (i.e., session-clusters) throughout the day, which is an important characteristic of mobile phone use that has not been reported in the existing literature.

Table 4. Average Rate of Repeated Sessions (RRS)*

|  | All Clusters | Large Clusters (including 10+ sessions) |
|---|---|---|
| N of Users | 4,017 | 3,966 |
| N of Clusters | 51,636 | 38,999 |
| Mean RRS | 48% | 60% |
| Median RRS | 48% | 63% |

* The unit of analysis is user-cluster-session.

Table 4 shows that the average RRS over all 4,017 users is 48% (mean and median), which means that nearly half of mobile phone tasks happen at a fixed time slot. A further analysis reveals that approximately one-fourth of the clusters have only 9 or fewer sessions. If the small clusters are excluded, the RRS will become higher (mean = 60% and median = 63%). Nearly two-thirds of mobile phone use appears to be *pre-scheduled* (i.e., predictably recurring) rather than impulsive or random.

*Profiling User Communities*

We use Louvain algorithm to identify the 13 communities of users sharing similar temporal patterns within each community. Table 5 summarizes the key characteristics of the communities.

Table 5. Descriptive Statistics of Session-induced Communities of Users

| Community ID | Size (N of Users) | Share (%) of the Sample | Key Characteristics |
|---|---|---|---|
| 1 | 356 | 9 | Obsessive, intense usage from 10:30 to midnight, spikes around 17:15 to 18:00, and 20:30 to 23:00 |
| 2 | 754 | 19 | Fourth quarter, starting at 18:00 and reach peak from 21:00 to 23:00 |
| 3 | 193 | 5 | Crescendo, repeated heavier usage with intervals of 2 or 3 hours, peaking at 18:00 to 19:00 |
| 4 | 338 | 8 | Whole evening, increased usage from 12:00 and stay highest from 17:00 to 22:00 |
| 5 | 78 | 2 | Mild, average usage across the day with a few heavier sessions in the evening |
| 6 | 289 | 7 | Obsessive, intense usage from 10:30 to midnight, spikes around 17:15 to 18:45, and from 21:00 to 22:30 |
| 7 | 634 | 16 | Fourth quarter, starting at 18:00 and reach peak 22:00 to 23:00 |
| 8 | 457 | 11 | Day leaper, spike starting at 22:00 till midnight |
| 9 | 199 | 5 | Second half, increasing usage from noon till 22:00, peak at 15:00 |
| 10 | 288 | 7 | Second half, increasing usage from noon till 22:00, peak at 16:30 |
| 11 | 420 | 10 | Obsessive, intense usage from 10:30 to midnight, spikes around 16:30 to 18:00, and 20:00 to 22:30 |
| 12 | 2 | 0.05 | Casual, short-spanned, sparse, heavy around 18:00 to 19:00 |
| 13 | 2 | 0.05 | Early starter, low during day, heavy around 6:30 and 19:30 |

To help interpret the user communities, we create a heat map to highlight the temporal patterns of each community throughout a 24-hour cycle. Fig. 5 shows a common temporal pattern for all user communities that has been observed elsewhere (e.g., Falaki et al., 2010; Van Canneyt et al., 2017), i.e., mobile phone use being moderately active during day time and intensified at evening, but being extremely rare in night.

Nevertheless, Fig. 5 reveals noticeable differences across communities that may not be noticeable otherwise under other ways of scrutiny. For instance, while 10 out of the 11 communities are active during the evening, they have different peak times . #9–#11 tend to be extremely busy around 18:00 (perhaps to schedule their dinner or social activities at night?) but slow down for roughly 2 hours (to enjoy their dinner or gatherings?) before intensifying again. #8 seems to dine with their phones all the time. #2–#6 peak between 21:00 and midnight, each starting and ending at approximately 30 minutes apart. The only exception (#1) is that the phone use intensifies around 17:00 (to get ready to go home?).

Figure 5.  Heatmap of Mobile Sessions by User-Communities*

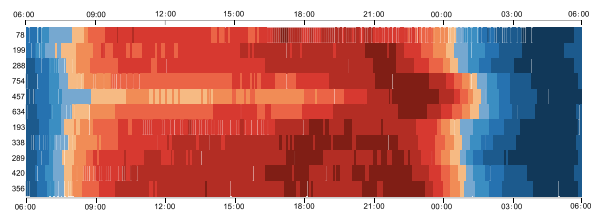

* Each row represents a user-community, labeled by the size of respective membership shown on the Y-axis.  Red color indicates an intensified level of activities, yellow color a moderate level and blue color a low level.  Two small communities, with 2 users each, are excluded.

**Conclusion and Discussions**

In this study, we demonstrate that sessions can serve an elementary measure of mobile phone use based on which the coherent session-clusters for each user and distinct user-communities can be formed successively to uncover temporal patterns within and across users.  By using the three cases, we also demonstrate with three cases how to apply the session-based measures, including sessions, session-clusters, and user-communities, to study theoretical or practical research questions.

The session-based measures are useful for many other ways unexplored in the current study.  One promising area is the application of the sequential modeling (Abbott, 1995) to session-based measures to examine how users select and organize different app use into a session and arrange various sessions in certain orders to satisfy different types of needs in daily life.  The sequence-level analysis of mobile phone use will provide an additional level of information about behavioral patterns, a level that is inaccessible with traditional equal-length measures (Bakeman & Gottman, 1986).

To help explain session, we contrast it with the conventional equal-length measures (e.g., the amount of time or duration).  However, session is not a replacement but rather a complement of equal-length measures.  In the literature of media use, McLeod and McDonald (1985) once argued that attention to media is a better measure of the audience behavior than exposure to media.  Subsequent research has commonly measured both exposure and attention simultaneously.  Therefore, we hope that future research on mobile phone use will also integrate equal-length and variable-length measures to maximize the advantages of the two.

This study has several limitations.  First, the Cambridge data is a convenience sample of volunteers.  Although methodological studies, such as ours, usually do not require a probability sample, it is necessary to note that the results of this study (e.g., the number and length of sessions) should be interpreted with caution.  Second, limited by the insufficient information about apps, we have constructed only multi-app sessions, which are less informative than single-app sessions.  Third, we have relied on unsupervised learning (i.e., community detection) to uncover temporal patterns.  As such, we will reiterate our caution when interpreting the results of the study.

In future research, we call for replications of the current study that will use probability samples and single-app sessions. We also invite new analytical frameworks and tools, such as deep learning methods to mine the rich information embedded in sessions (LeCun, Bengio & Hinton, 2015).  Finally, our session-based approach can be integrated into a variety of substantive research on the use of mobile phones and beyond such as user profiling, consumer lifestyle, spatial mobility, and social movements.

**Appendix 1. Pseudo Code for Identifying and Constructing Sessions**

```
screenon = False
screenoff = False
keyguardoff = False
for line in open(log):
   if not screenon:
     screenon =
     line.contains('screen_on')
     continue
   if not keyguardoff:
     keyguardoff = line.contains
        ('keyguard_removed')
     if not keyguardoff:
        screenon = False
     else:
        screen_unlocked = line
     continue
   if not screenoff:
     screenoff = line.contains
        ('screen_off' OR 'shutdown'))
     if screenoff:
        screen_locked = line
        write(screen_unlocked)
        write(screen_locked)screenon =
          False
           screenoff = False
        keyguardoff = False
```

**Appendix 2. Aggregation of Sessions based on Community Detection**

In community detection, "similar" nodes of a network are grouped into the same community (or cluster) such that the nodes are as homogenous as possible within the same community but as heterogeneous as possible to all other communities. The idea can be applied to other situations that may not look like a network but do exist an implicit network structure. In the current study, we consider each session to be a node, and the temporal similarity between any pair of sessions an edge. As such, all sessions of a user form a temporal network. We use the inverse of a 2-dimensional Euclidian distance ($d_{i,pq}$) to quantify the edge between sessions $p$ and $q$ ($p \neq q$) for each user $i$ (= 1 to $n$):

$$d_{i,pq} = 1/\sqrt{\sum_{p,q=1, p \neq q}^{m}(S_{i,pt} - S_{i,qt})^2 + (E_{i,pt} - E_{i,qt})^2}, \quad (A1)$$

where $S_{i,pt}$ and $S_{i,qt}$ are the start time of $p$ and $q$, $E_{i,pt}$ and $E_{i,qt}$ are the end time of $p$ and $q$, respectively, with $p$ and $q$ varying from 1 to $m$ per $i$, and $m$ is the total number of sessions for $i$. Note that the above summation does not involve $i$, implying that Eq. A1 is applied to each user separately (i.e., each user is a network).

In calculating the Euclidean distances, we consider only the time portion (i.e., hour, minute, and second) of the timestamp for all sessions while discarding the date portion (i.e., year, month, and day), based on the findings that media use behavior follows a *cyclical*-time system (i.e., repeated over 24 hours), instead of a *linear*-time system (forwarding from one day to next) (e.g., Manley, Zhong & Batty, 2016; Van Canneyt, Bron & Haines, 2017). Consequently, the distance between sessions is determined only by the clock, not by the calendar. Consider three sessions A, B, and C of a user. A occurred at 0:01 on day 1, B at 23:59 the same day, and C at 23:59 next day. The distance among the three sessions would be quite large (almost 24 or 48 hours) based on a normal linear timeline; but was very small (1 or 2 minutes) based on a clock because they all took place around the midnight, which fits the essence of temporal similarity.

We also apply the above procedure to detect temporal similarity among users based on their session patterns, by making minor changes to Eq. A1:

$$d_{i,uv} = \sqrt{\sum_{i=1}^{n}\sum_{u,v=1,u \neq v}^{m}(C_{iu} - C_{iv})^2}, \quad (A2)$$

where $C_{iu}$ and $C_{iv}$ are the centroid of session-clusters $u$ and $v$ ($u \neq v$), respectively, with $u$ and $v$ varying from 1 to $m$ for all $i$'s, and $m$ refers to the total number of session-clusters for $i$. Both $C_{iu}$ and $C_{iv}$ are given by their most "central" member sessions (based on closeness centrality), respectively. To distinguish the two rounds of community detection, we call the resulting communities from Eq. A1 as "session-*clusters*" and the communities from Eq. A2 as "user-*communities*".